\documentclass[%
 reprint,
 amsmath,amssymb,
 aps, prl,
 longbibliography,
 twocolumn
]{revtex4-1}

\usepackage{graphicx}
\usepackage{dcolumn}
\usepackage{bm}
\usepackage{siunitx}
\usepackage{nicefrac}

\usepackage[plainpages=false,pdfpagelabels,colorlinks=true,linkcolor=black,urlcolor=black,citecolor=black]{hyperref}

\begin{document}

\renewcommand{\vec}[1]{\mathbf{#1}}

\title{Suppressing dissipation in a Floquet-Hubbard system}

\author{Konrad Viebahn}
\affiliation{Institute for Quantum Electronics, ETH Zurich, 8093 Zurich, Switzerland}
\author{Joaqu\'in Minguzzi}
\affiliation{Institute for Quantum Electronics, ETH Zurich, 8093 Zurich, Switzerland}
\author{Kilian Sandholzer}
\affiliation{Institute for Quantum Electronics, ETH Zurich, 8093 Zurich, Switzerland}
\author{Anne-Sophie Walter}
\affiliation{Institute for Quantum Electronics, ETH Zurich, 8093 Zurich, Switzerland}
\author{Frederik G\"org}
\affiliation{Institute for Quantum Electronics, ETH Zurich, 8093 Zurich, Switzerland}
\author{Tilman Esslinger}
\email{esslinger@phys.ethz.ch}
\affiliation{Institute for Quantum Electronics, ETH Zurich, 8093 Zurich, Switzerland}

\date{\today}

\begin{abstract}
The concept of `Floquet engineering' relies on an external periodic drive to realise novel, effectively static Hamiltonians.
This technique is being explored in experimental platforms across physics, including ultracold atoms, laser-driven electron systems, nuclear magnetic resonance, and trapped ions.
The key challenge in Floquet engineering is to avoid the uncontrolled absorption of photons from the drive, especially in interacting systems in which the excitation spectrum becomes effectively dense.
The resulting dissipative coupling to higher-lying modes, such as the excited bands of an optical lattice, has been explored in recent experimental and theoretical works,
but the demonstration of a broadly applicable method to mitigate this effect is lacking.
Here, we show how two-path quantum interference, applied to strongly-correlated fermions in a driven optical lattice, suppresses dissipative coupling to higher bands and increases the lifetime of double occupancies and spin-correlations by two to three orders of magnitude.
Interference is achieved by introducing a weak second modulation at twice the fundamental driving frequency with a definite relative phase.
This technique is shown to suppress dissipation in both weakly and strongly interacting regimes of a driven Hubbard system, opening an avenue to realising low-temperature phases of matter in interacting Floquet systems.
\end{abstract}

\maketitle

Dissipation emerges when a system is coupled to a large number of degrees of freedom in its environment.
In periodically driven systems dissipation thus naturally arises when the low-energy modes are coupled to lossy excited modes by the drive.
This presents a formidable challenge to Floquet engineering, in which periodic driving is used to create a host of novel, effectively static Hamiltonians, with ultracold atoms~\cite{goldman_periodically_2014, bukov_universal_2015, holthaus_floquet_2016, eckardt_colloquium:_2017} and beyond~\cite{oka_floquet_2019, harper_topology_2020, peng_observation_2019, kiefer_floquet-engineered_2019}.
The success of Floquet engineering relies on the existence of a favourable timescale on which a given Floquet Hamiltonian remains valid before heating kicks in and harms the quantum state.
Adding interactions to the picture, the effects of unwanted energy absorption is further complicated by a dense excitation spectrum~\cite{reitter_interaction_2017, singh_quantifying_2019, boulier_parametric_2019, wintersperger_parametric_2020, rubio-abadal_floquet_2020, kuwahara_floquetmagnus_2016, moessner_equilibration_2017}.
Consequently, the choice of Floquet driving frequency is always a compromise~\cite{sun_optimal_2020}.
On the one hand, the Floquet driving frequency should be chosen as high as possible so as to maximise the detuning between the drive and the natural (low-)energy scale of interest~\cite{sun_optimal_2020, peng_observation_2019, rubio-abadal_floquet_2020}.
On the other hand, the presence of higher-lying modes of the underlying Hamiltonian, such as energetically higher Bloch bands in an optical lattice~\cite{bakr_orbital_2011, weinberg_multiphoton_2015, flaschner_high-precision_2018, messer_floquet_2018, cabrera-gutierrez_resonant_2019, holthaus_floquet_2016, eckardt_colloquium:_2017, strater_interband_2016, keles_effective_2017, quelle_resonances_2017, sun_optimal_2020}, pose a limit on how high the drive frequency can be.
In practice, the dissipative coupling to higher bands, due to tunnelling within the excited band, inhomogeneities, or subsequent excitations to even higher bands, are the limiting factor in many Floquet driving schemes.
While tailored lattice potentials can be employed to push excited bands to higher frequencies and keeping the lowest band dispersive~\cite{messer_floquet_2018}, a broadly applicable solution to counteract this form of dissipation has so far not been demonstrated in many-body systems in which the excited modes are densely spaced.

\begin{figure*}[t!]
\begin{center}
\includegraphics[width = 1.0\textwidth]{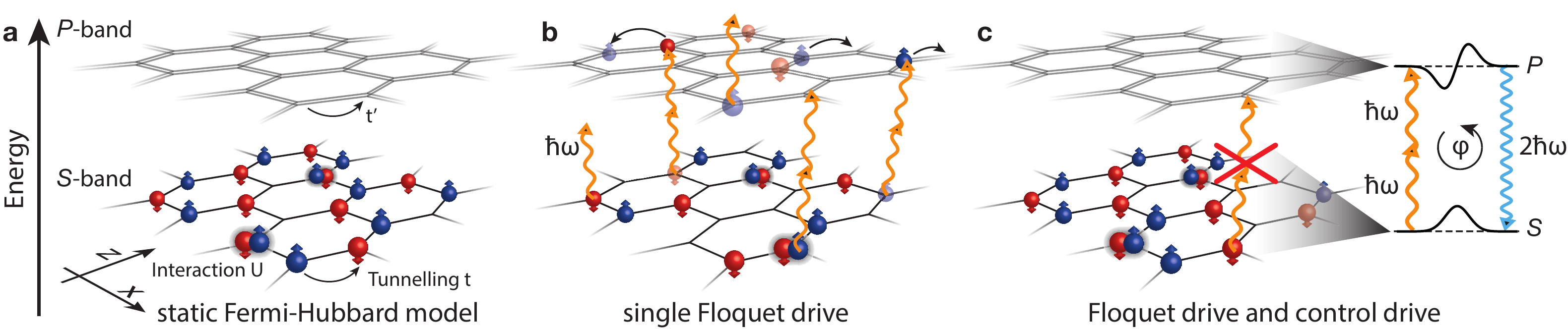}
\caption{
  Two-frequency driving suppresses dissipative coupling to higher bands.
  (a) Ultracold fermionic atoms in two internal states $\uparrow,\downarrow$ occupy a three-dimensional, hexagonal optical lattice ($y$-direction omitted for clarity).
  They are described by the ground-band Fermi-Hubbard model with characteristic energies $t$ (tunnelling) and $U$ (on-site interaction), while energetically higher-lying bands (tunnelling $t'$) are not populated in the absence of periodic driving.
  Here, only the fifth band is shown, which originates from hybridised ($P$-)orbitals in $x$-direction. 
  (b) Periodic driving close to resonance with an inter-band transition leads to dissipation in the ground-band Fermi-Hubbard system, originating from excited-band tunnelling ($t'$), inhomogeneities, or subsequent excitation to even higher bands.
  The first relevant resonance is a two-photon process at frequency $\omega$ (shown in orange), whereas the single-photon transition is located at $2\omega$ (blue).
  (c) When applying two driving frequencies simultaneously with matched transition strengths, tuning the relative phase $\varphi$ coherently enhances or suppresses dissipative coupling to higher bands.
}
\label{fig:1}
\end{center}
\end{figure*}

In this paper, we study strongly-correlated fermions in an amplitude-modulated optical lattice, i.e.~the driven Fermi-Hubbard model, in which the drive couples dissipatively to higher Bloch bands.
We demonstrate control of dissipation by introducing a second excitation pathway at twice the fundamental driving frequency and tuning the relative phase~\cite{schiavoni_phase_2003,
zhuang_coherent_2013, niu_excitation_2015,
gorg_realization_2019} between the two drives to maximise quantum interference.

The starting point for our experiments is an equal spin mixture of $45'000$ to $60'000$ ultracold potassium-40 atoms loaded in an optical lattice, realising the Fermi-Hubbard model (Fig.~1a).
The three-dimensional, hexagonal lattice structure (the $y$-direction is not shown in Fig.~1a) is characterised by tunnel couplings 
$(t_x, t_y, t_z)/h = (340, 90, 106)$ Hz~\footnote{See Supplemental Material, which includes Refs.~\cite{greif_short-range_2013, uehlinger_artificial_2013}, for further experimental details, theoretical calculations, and data analysis.} and the Hubbard $U/h$ can be tuned to values between \SI{-7}{kHz} and \SI{+7}{kHz} ($h$ is Planck's constant).
\nocite{greif_short-range_2013, uehlinger_artificial_2013}
Of the lowest two ($S$-type) bands of the optical lattice the ground band is typically populated by $65\%$ of the atoms.
The excited-orbital bands ($P$, $D$, \dots) give rise to a multi-band Hubbard model and are not populated in the static case (Fig.~1a).

Periodic driving is introduced via amplitude modulation of the lattice depth in $x$-direction with the waveform
\begin{equation}
  V_{\bar{\text{X}}}(\tau) = V_0\times\left[1 + A_{\omega}\cos(\omega \tau) + A_{2\omega}\cos(2\omega \tau + \varphi)\right]\quad\text{,}
\end{equation}
leading, predominantly, to time-periodic modulation of the tunnelling energy $t_x$~\cite{Note1}.
In the Floquet framework, the periodic modulation of $t_x(\tau)$ could be re-cast into an effective (static) tunnelling $t_{\text{eff}}$ when modulating at resonance with an energy scale of the underlying Hamiltonian~\cite{bukov_universal_2015, eckardt_colloquium:_2017}.
For instance, modulating $t$ resonantly with both a static site-offset and the Hubbard $U$ allows for engineering of anyon-Hubbard models~\cite{cardarelli_engineering_2016}.
In general, however, lattice modulation also causes unwanted higher-band excitations, particularly for the large driving amplitudes typically required for Floquet engineering~\cite{bukov_universal_2015, eckardt_colloquium:_2017, cardarelli_engineering_2016}.
In particular, the appearance of multi-photon resonances (Fig.~1b) represents a limitation on the choice of Floquet driving frequency and amplitude~\cite{weinberg_multiphoton_2015, strater_interband_2016, sun_optimal_2020}.

We characterise single- and multi-photon resonances experimentally by tuning the Hubbard $U$ to zero and recording the excitation spectrum for two different values of the single-frequency modulation amplitude $A_{\omega}$ (Fig.~2a, $A_{2\omega} = 0$).
Here, we modulate the lattice at frequency $\omega$ for a duration of \SI{20}{ms} and record the number of atoms in the ground band ($N_{\text{ground}}$) by counting atoms in Brillouin-zones after band-mapping and time-of-flight~\cite{Note1}. 
The measurement is in good agreement with a numerical solution to the time-dependent Schr{\"o}dinger equation, taking into account the non-separable lattice potential in the $xz$-plane but neglecting any inhomogeneities~\cite{Note1}.
Since the modulation is performed on the $t_x$-bonds, atoms are predominantly excited to the fifth ($P$-)band which arises from hybridised orbitals in $x$-direction.
Correspondingly, the spectrum in Figure~2a shows pronounced dips in the ground-band population at \SI{24.5}{kHz} (single-photon resonance) as well as multi-photon resonances at \SI{12.25}{kHz} (two-photon) and \SI{8.17}{kHz} (three-photon).
When modulating for a varying amount of time, we never observe a revival of the ground band population, moreover, the excitations are accompanied by severe atom loss during the modulation (see below).
These observations confirm the dissipative nature of the resonance features.

\begin{figure}[htbp]
\begin{center}
\includegraphics[width = 0.48\textwidth]{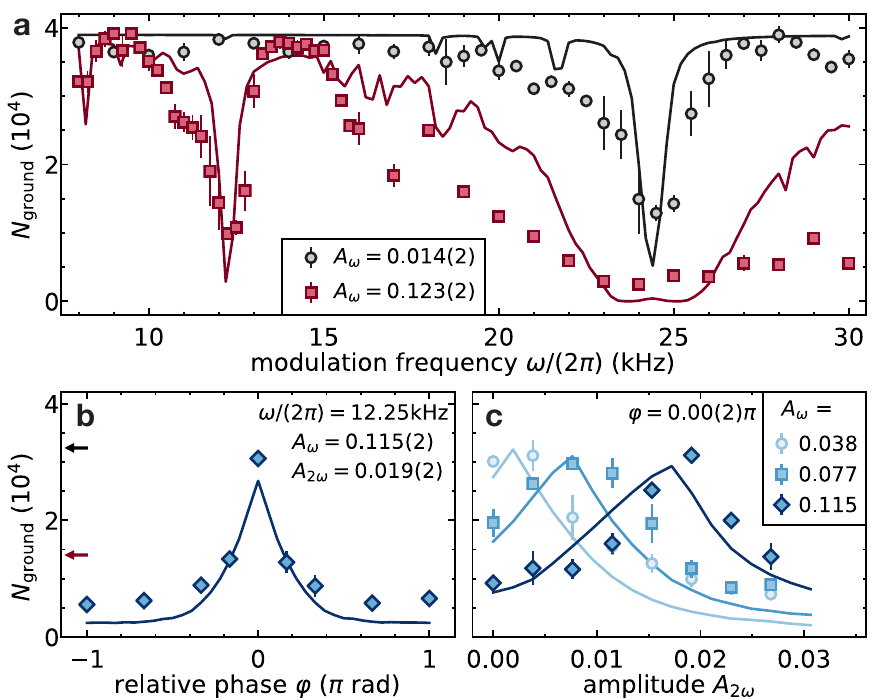}
\caption{
  Optimising dissipation control in the non-interacting regime.
  (a) Single-frequency excitation spectrum for non-interacting atoms.
  The data was obtained by modulating the amplitude of the lattice beam in $x$-direction for \SI{20}{ms} for two relative strengths $A_{\omega}$ of 0.014(2) (grey circles) and 0.123(2) (red squares).
  Single- and multi-photon resonance features appear at the $S$-$P$ inter-band transition at \SI{24.5}{kHz} and fractions thereof, in agreement with the theoretical calculation (lines) that takes into account the full lattice potential in the $xz$--plane~\cite{Note1}.
  (b) Ground band population $N_{\text{ground}}$ after \SI{20}{ms} of two-frequency modulation (Eq.~1) as function of relative phase $\varphi$.
  A matched excitation strength gives rise to a strong dependence on relative phase in agreement with ab-initio theory (lines).
  The black (red) arrow shows the ground band population in the absence of any drive (in the presence of a single drive at $\omega$).
  (c) Ground band population at $\varphi = 0$ as function of control amplitude $A_{2\omega}$ for various Floquet amplitudes $A_{\omega}$.
  As the Floquet amplitude increases, the optimum control amplitude is shifted to larger values.
  Error bars to measured data points are the standard error of at least three measurements~\cite{Note1}.
}
\label{fig:1}
\end{center}
\end{figure}

In the following we outline the strategy to mitigate the previously described dissipation process via destructive interference.
Modulation at frequency $\omega$ (let us call it the `Floquet' drive) effectively leads to a Rabi-coupling to the higher band~\cite{bakr_orbital_2011}, which in general can be written as a complex number.
Now, the key idea is to add a second modulation at $2\omega$ (the `control' drive) that produces a Rabi-coupling of the same magnitude as the Floquet drive but with opposite sign, such that both couplings add up to zero, i.e.~interfere destructively.
Experimentally, we introduce the control drive with amplitude $A_{2\omega}$ and vary the relative phase $\varphi$ between the two drives (Fig.~2b, $A_{\omega} = 0.115(2)$, $A_{2\omega} = 0.019(2)$).
In the presence of the control drive, the ground band atom number $N_{\text{ground}}$ shows a strong dependence on $\varphi$, which is direct evidence of phase-only control of the dissipation channel.
In particular, $N_{\text{ground}}$ peaks at $\varphi = 0.00(1)\pi$, corresponding to a time-reversal-symmetric waveform (Eq.~1) for which both couplings are real-valued but differ by a relative minus sign.
This measurement is in good agreement with the theoretical prediction (lines in Fig.~2), as maximal destructive interference leads to a ground band atom number as high as 30.6(8)$\times 10^3$, which corresponds to $94\%$ of the static value (black arrow in Fig.~2b).
In contrast, a relative phase of $\pi$ causes depopulation of the ground band with less than $20\%$ of atoms remaining, compared to $43\%$ in the absence of the control drive (red arrow).
Fixing the relative phase to $\varphi = 0$ allows us to achieve destructive interference for increasing values of Floquet amplitude $A_{\omega}$ with the optimum control amplitude shifting to larger values, in good agreement with theory, as is shown in Figure~2c.
This observation highlights the flexibility of the dissipation control scheme which is not bound to any particular Floquet amplitude.
As the single-photon excitation strength is much larger than the two-photon coupling~\cite{strater_interband_2016}, only a weak control amplitude is necessary to achieve control.

\begin{figure}[htbp]
\begin{center}
  \includegraphics[width = 0.48\textwidth]{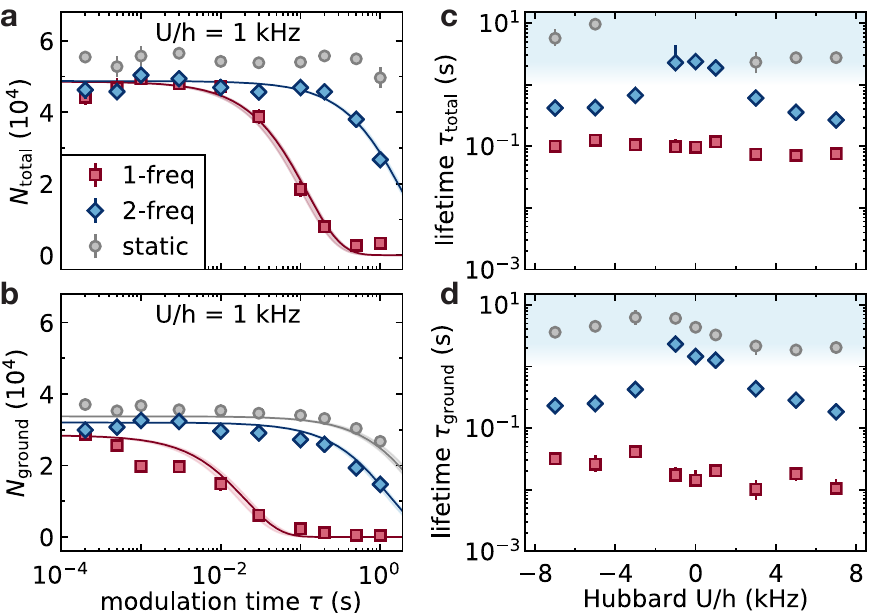}
  \caption{Effect of interactions on suppressing dissipation.
  Measured total atom number $N_{\text{total}}$ (a) and ground band atom number $N_{\text{ground}}$ (b) as function of modulation time, to which an exponential decay is fitted (lines).
  While a single-frequency drive leads to strong atom loss and de-population of the ground band via higher-band excitations (red points), the presence of the control drive at $2\omega$ (blue) cancels this dissipation mechanism.
  (c-d) Resulting $1/e$-lifetimes of the total atom number ($\tau_{\text{total}}$, c) and ground band population ($\tau_{\text{ground}}$, d) as function of Hubbard $U$.
  Since the longest measurement time is \SI{1}{s}, fitted lifetimes longer than this value can be unreliable (blue shaded area).
  Data points in (a-b) show the mean and standard error of three individual measurements, whereas the points in (c-d) result from a least-squares fit.
  The shaded areas in (a-b) and the error bars in (c-d) show the estimated uncertainty of the fitted value via bootstrapping~\cite{Note1}.
}
\label{fig:3}
\end{center}
\end{figure}

Having verified the control technique in the single-particle case, we now investigate the influence of interactions on cancelling the dissipative coupling to higher bands.
The many-body state in the driven Fermi-Hubbard model is experimentally characterised by four different observables, namely total atom number $N_{\text{total}}$, ground band population $N_{\text{ground}}$, double occupancy, and nearest-neighbour spin correlations~\cite{Note1}.
In order to maximise our sensitivity, we choose to drive our system resonantly with a two-photon higher-band transition with $\omega/(2\pi) = \SI{12.25}{kHz}$ with an amplitude of $A_{\omega} = 0.115(2)$, giving rise to a strong dissipative coupling.
The driving frequency is chosen to be off-resonant to any static energy scale such that the lowest-band effective Floquet Hamiltonian is equal to the Fermi-Hubbard model, i.e.~$t_{\text{eff}} = t_x$.
In particular, the Floquet frequency is larger than all probed values of Hubbard $U$, thus precluding density-assisted tunnelling and creation of double occupancies~\cite{messer_floquet_2018}.
In the experiment, we modulate the system for a varying duration $\tau$ and different values of Hubbard $\vert U\vert<\omega$.
This data is then fitted by an exponential decay in order to extract a lifetime.
For each measurement we compare single-frequency modulation ($A_{\omega} = 0.115(2)$, $A_{2\omega} = 0$) to the two-frequency driving scheme ($A_{\omega} = 0.115(2)$, $A_{\omega} = 0.019(2)$, $\varphi = 0.00(1)\pi$), as well as the static system.

The measured lifetime of both the total atom number $\tau_{\text{total}}$ and the ground band population $\tau_{\text{ground}}$ (Fig.~3) show a strongly increased lifetime in the presence of the control drive, compared to the singly-driven case. 
For weak and intermediate interactions ($\vert U\vert/h \lesssim$ \SI{1.5}{kHz} $ =$ bandwidth), the lifetime of the ground band atom number (Fig.~3d) is increased by more than two orders of magnitude from $20\substack{+3 \\ -3}\,$ms to $1.3\substack{+0.1\\-0.1}\;$s at $U/h=\SI{+1}{kHz}$, comparable to the static lifetime.
While for strong interactions ($U \gtrsim$ bandwidth) the lifetimes in the two-frequency driving protocol are reduced, compared to the weakly interacting case, applying dissipation control still increases the lifetime by about an order of magnitude.
For example, the lifetime of the ground band atom number at strong repulsive interactions ($U/h = \SI{+7}{kHz}$) is increased eighteen-fold from $10\substack{+4 \\ -2}\,$ms to $184\substack{+27 \\ -9}\,$ms, corresponding to more than 2000 Floquet driving cycles.

\begin{figure}[tbp]
\begin{center}
  \includegraphics[width = 0.48\textwidth]{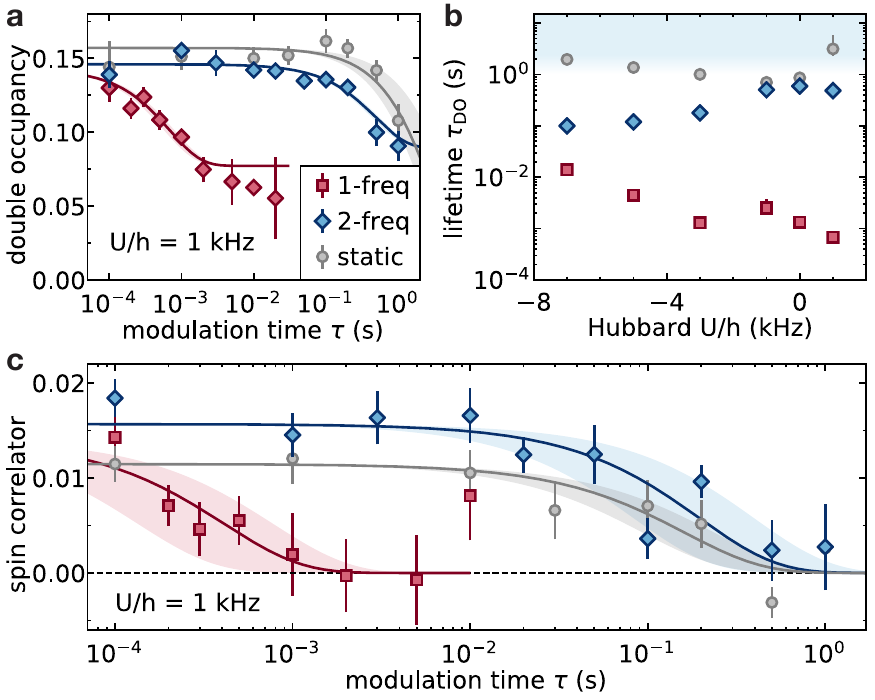}
  \caption{
    Suppressing dissipation in the strongly-correlated Fermi-Hubbard model.
    (a) Measured double occupancy as function of modulation time. Double-occupancy is the fraction of atoms in doubly-occupied lattice sites in the ground band.
    When applying two-frequency control (blue), the level of double occupancy remains almost unchanged up to \SI{1}{s}, similar to the static case (grey).
    (b) Fitted exponential lifetimes of double occupancy $\tau_{\text{DO}}$ as function of Hubbard $U$. For weak and intermediate interactions, the control method (blue) restores the lifetime of double occupancy almost to the static value (grey).
    We restrict this measurement to values of Hubbard $U/h \leq \SI{+1}{kHz}$ as the initial value of double occupancy for strong repulsive interactions becomes too low to reliably extract a lifetime.
    (c) Tracing the nearest-neighbour spin correlator~\cite{Note1} with modulation time for $U/h=\SI{1}{kHz}$ in the presence of the control drive (blue), compared to the single-frequency case (red).
    The static reference values are plotted in grey.
  Data points in (a) and (c) show the mean and standard error of at least three and ten individual measurements, respectively, whereas the plotted lifetimes in (b) result from a least-squares fit.
  The shaded areas in (a,c) and the error bars in (b) show the estimated uncertainty of the fitted value via bootstrapping~\cite{Note1}.
  Since the longest measurement time is \SI{1}{s}, fitted lifetimes longer than this value can be unreliable (blue shaded area in b).
}
\label{fig:4}
\end{center}
\end{figure}

The improved lifetimes in the band-mapping measurements (Fig.~3) strongly suggest that dissipative coupling to higher bands can be efficiently counteracted by adding the control drive, even for large Hubbard $U$.
However, the single-frequency drive already leads to heating on timescales shorter than the onset of atoms loss in the ground band (10 to \SI{100}{ms}).
In order to investigate this effect, we complement our previous data with measurements of ground-band double occupancy, that is, the fraction of atoms in doubly-occupied sites in the lowest Bloch band~\cite{messer_floquet_2018}.
This observable is ideally suited to characterise many-body states in the driven Fermi-Hubbard model and it has been used in the past for an ab-initio comparison of our experiment with non-equilibrium dynamical mean-field theory~\cite{sandholzer_quantum_2019}.
In the experiment, we start with a certain level of double occupancy in the static lattice set by the value of $U$, assuming a thermally equilibrated state~\cite{messer_floquet_2018, sandholzer_quantum_2019}.
As before, we introduce the modulation for varying amounts of time $\tau$ and then freeze the lattice in order to suppress any further dynamics~\cite{Note1}.
Atoms on doubly-occupied sites are then associated to Feshbach molecules which experience a shifted zero-point on-site energy.
As this on-site energy depends on the Bloch band, we are able to spectroscopically resolve double-occupancies in the ground band; atoms in higher bands are expelled from the trap during the band-mapping ramp while atoms in the lowest band remain trapped~\cite{Note1}.
The results of this measurement and the extracted lifetimes ($\tau_{\text{DO}}$) are shown in Figure~4a-b.
We observe a fast drop in double occupancy ($\tau_{\text{DO}} = 0.7\substack{+0.1\\-0.1}$ms for $U/h = \SI{+1}{kHz}$) in the single-frequency case, which precedes the ground band atom loss ($\tau_{\text{DO}}\ll\tau_{\text{ground}} = 20\substack{+3\\-3}$ms, Fig.~3d).
This observation is consistent with a fast increase in entropy, i.e.~heating, which causes the cloud to expand in the trap, resulting in a lower density at the centre of the trap and reduced double occupancy.
In stark contrast, the lifetime of double occupancy is increased by up to a factor of hundred when applying two-frequency control.
For example, this lifetime exceeds \SI{1}{s} for $U/h=\SI{+1}{kHz}$ which is similar to our static lifetimes, reaching to the limit within which we can reliably extract a lifetime (blue shaded areas).
This measurement corroborates the previous results for strongly attractive and weakly repulsive interactions (Fig.~3c-d).

While charge dynamics (e.g., double occupancy) in the Fermi-Hubbard model are governed by $t$ and $U$, the lowest hierarchy in the system is the spin exchange energy $4t^2/U$.
Therefore, spin-correlations are highly susceptible to temperature changes and we thus expect them to vanish quickly in the single-frequency driven case.
The experimental data is plotted in Figure~4c which shows that nearest-neighbour spin correlations~\cite{Note1} are indeed destroyed within just a few driving cycles, with a $1/e$-lifetime of $410\substack{+360\\-280}\;\mu$s.
Remarkably, dissipation control extends the lifetime of nearest-neighbour spin-correlations by three orders of magnitude, to $200\substack{+200 \\ -110}$\;ms which is within measurement error of the static value ($160\substack{+80 \\ -70}$\;ms).

In conclusion, we demonstrated how adding a weak harmonic to the fundamental driving frequency can suppress dissipative coupling to higher-lying modes in a driven many-body system, even in the presence of strong interactions.
Long lifetimes in strongly-driven, ultracold matter are a prerequisite for realising low-temperature phases in these systems, for example in an anyon-Hubbard model~\cite{cardarelli_engineering_2016}.
A natural extension of this scheme employs additional control frequencies to cancel further resonances, in the spirit of pulse-shaping techniques in laser-driven molecular dynamics~\cite{kuroda_mapping_2009}.
Moreover, tuneable couplings to higher bands can be an asset, rather than a nuisance, for instance, in designing novel cooling schemes, not only for optical lattices~\cite{bakr_orbital_2011, arnal_evidence_2019} but also in condensed matter~\cite{werner_light-induced_2019}.
More generally, this kind of precisely engineered dissipation is a valuable resource for engineering specific quantum states, particularly in the many-body context~\cite{muller_engineered_2012}.

\appendix

\section{Acknowledgments}

\begin{acknowledgments}
  We would like to thank J.~L{\'e}onard, D.~Malz, M.~Messer, Y.~Murakami, P.~Werner, and W. Zwerger for discussions and comments on the manuscript. 
This work was partly funded by the SNF (project nos.~169320 and 182650), NCCR-QSIT, QUIC (Swiss State Secretary for Education, Research and Innovation contract no.~15.0019), and ERC advanced grant TransQ (project no.~742579).
K.V.~is supported by the ETH Zurich Postdoctoral Fellowship programme.
\end{acknowledgments}


%

\newpage
\cleardoublepage

\setcounter{figure}{0} 
\setcounter{equation}{0} 

\renewcommand\theequation{S\arabic{equation}} 
\renewcommand\thefigure{S\arabic{figure}} 
\setcounter{page}{1}

\section{Supplemental Material}

\subsection{General preparation}
To realize the driven Fermi-Hubbard model, we first prepare a gas of $^{40}$K fermionic atoms in the two magnetic sublevels $m_F=-9/2,-7/2$ of the $F=9/2$ manifold, which is trapped in a harmonic optical dipole trap. We prepare an incoherent spin-balanced mixture of the two spins by performing many radio-frequency sweeps over the transition resonance between the two states. The cloud is then evaporatively cooled to quantum degeneracy at a scattering length of $a = 116(1)a_0$, where $a_0$ is the Bohr radius. The mean number of atoms in this cloud is $46(2)\cdot 10^3$ at a temperature of $0.09(1)\nicefrac{T}{T_F}$. For strong attractive and weak repulsive interactions we use the $-9/2,-7/2$ mixture, whereas for strong repulsive interactions we use a $-9/2,-5/2$ mixture. The latter is prepared by applying a Landau-Zener sweep to transfer all atoms from the $-7/2$ to the $-5/2$ spin state.

We create a three-dimensional optical lattice by retro-reflecting four beams of wavelength $\lambda = 1064 \, \text{nm}$. The resulting lattice potential perceived by the atoms is:
\begin{eqnarray} V(x,y,z) & = & -V_{\overline{X}}\cos^2(k_L
x+\theta/2)-V_{X} \cos^2(k_L
x)\nonumber\\
&&-V_{\widetilde{Y}} \cos^2(k_L y) -V_{Z} \cos^2(k_L z) \\
&&-2\alpha \sqrt{V_{X}V_{Z}}\cos(k_L x)\cos(k_Lz)\cos\varphi_{SL} \nonumber, 
\label{Lattice}
\end{eqnarray}
where $k_L=2\pi/\lambda$ and $x,y,z$ are the three experimental axes. The lattice depths $V_{\overline{X},X,\tilde{Y},Z}$ are measured in units of the recoil energy $E_R=h^2/2m\lambda^2$, where $h$ is the Planck constant and $m$ the mass of the atoms. Each lattice depth is independently calibrated by modulating its amplitude in the presence of a $^{40}$K degenerate Fermi cloud.
The visibility $\alpha=0.958(8)$ is also calibrated using amplitude modulation on a degenerate $^{40}$K Fermi cloud, but in an interfering lattice configuration. The phase $\theta$ that fixes the geometry of the lattice is set to $\theta=1.000(2)\pi$, whereas the superlattice phase $\varphi_{SL}$ is actively stabilized to $0.00(3)\pi$.

\subsection{Lattice configurations and modulation}

The potential from Eq.$~$\ref{Lattice} yields a hexagonal lattice structure, which is characterized by the lowest-band tunnelling rates $t_x$, $t_{x,\text{inter}}$, $t_y$ and $t_z$.
We load the atoms from the optical trap into an intermediate lattice within $\SI{200}{ms}$, where the tunnelling $t_{x,\text{inter}}$ between dimers remains finite $V_{\overline{X},X,\widetilde{Y},Z}/E_R=\left[7.89(9),0.10(2),9.7(2),8.2(2)\right]$.
We then linearly ramp up the lattice in $\SI{20}{ms}$ to a hexagonal configuration with $V_{\overline{X},X,\widetilde{Y},Z}/E_R=\left[21.0(3),3.70(4),9.7(2),6.81(8)\right]$ in which the modulation is performed.
Here, the resulting tunnelling energies are $\left[t_x, t_{x, \text{inter}}, t_y, t_z\right]/h = \left[340(50), 0.8(1), 90(4), 106(7)\right]$\;Hz, where the inter-dimer tunnelling $t_{x,\text{inter}}$ is negligibly small.\\
To implement the periodic drive in the system, we modulate the intensity of the non-interfering, $\overline{X}$  lattice beam by applying a sinusoidal signal to the RF power of an AOM resulting in a modulation of the lattice depth.
We obtain a periodically varying lattice depth of the form $V_{\overline{X}}\left(\tau\right)=V_{\overline{X}}\left(1+A_{\omega}\sin\left(\omega \tau\right)\right)$, where $\omega$ corresponds to the `Floquet' driving frequency and $A_{\omega}$ to the relative amplitude of the drive.
The lattice modulation leads to a modulation of all Hubbard parameters.
However, the influence of the modulation on $t_z$ remains smaller than $13\%$ and that on $U$ smaller than 5$\%$.
Such a periodic modulation will also introduce transitions to higher bands. To compensate for these losses, we add an additional `control' contribution to the drive $A_{2\omega}\sin\left(2\omega t+ \varphi\right)$ at twice the frequency.
The amplitude and phase of the drive are calibrated by recording the intensity of the lattice beam and performing a fast Fourier-transform on the signal. The remaining systematic deviations are $\SI{2}{\degree} = 0.01 \pi$ in the single-frequency driving phase and 0.002 in the amplitude; statistical deviations are smaller than these values.\\
In all measurements presented in the paper the amplitude of the modulation is ramped to its final value within two periods of the fundamental modulation frequency $\omega$ and suddenly switched off after the duration of the modulation $\tau$.

\subsection{Detection methods}

The experimental realization of the Fermi-Hubbard model is characterized by four observables: atom number and band population in Figure~3, as well as double occupancy fraction and spin correlations in Figure~4.\\

\textit{Band-mapping detection.} The atom number and band population are obtained through a detection method called band-mapping. Directly after the modulation we ramp down the optical lattice slowly enough for the atoms to stay adiabatically in their band but fast enough to avoid redistribution between bands. We thereby map quasi-momentum to real momentum. We do this by an exponential ramp to zero of the lattice depth within $\SI{500}{\micro s}$. After $\SI{500}{\micro s}$ we switch off the homogeneous magnetic field and trapping potential and allow for $\SI{15}{ms}$ time of flight (TOF) to map momentum onto position and thereby resolve the Brillouin zones in real space. We then take an absorption image of the expanded cloud.
 To assess the distribution of atoms among bands, we divide the image into zones for the different bands and integrate the atomic density in each zone. The centre position of the first Brillouin zone is determined by fitting a Gaussian to a $^{40}$K cloud released adiabatically from the trap. The size of the BZ is determined by a separate calibration method where we use a $^{87}$Rb condensate and flash the lattice: the $2\hbar k_L$ diffraction peaks yield the edges of the first zone and are corrected by a factor $\nicefrac{87}{40}$ due to the mass difference between $^{87}$Rb and $^{40}$K.\\ 
 
 \textit{Stern-Gerlach detection.} The measurement of double occupancy fractions and spin-correlations requires another detection method. 
 We release the atoms from the lattice within $\SI{100}{ms}$: longer than the $\SI{500}{\micro s}$ from the band-mapping to allow for the atoms to redistribute into Gaussian cloud shapes and for atoms in higher bands to be lost from the trap but still short enough to avoid atom loss.
 Also, these observables require a spin-resolved measurement: to distinguish between the different Zeeman sublevels we apply a short magnetic field gradient which leads to a separation of the spin states during TOF of $\SI{8}{ms}$.\\ 
 
\textit{Double-occupancy detection.} Directly after the modulation we freeze the dynamics through a quench to a deep cubic, 'freeze' lattice $V_{\overline{X},X,\widetilde{Y},Z}=\left[30.6(5),0.0,39.9(2),29.6(8)\right]$ within $\SI{100}{\micro s}$.
We then ramp the magnetic field over the $-\nicefrac{7}{2}$, $-\nicefrac{9}{2}$ Feshbach resonance at $B=\SI{202.1}{G}$ and spectroscopically resolve the interaction shift by radio-frequency radiation. As a result, only the atoms sitting in pairs on a lattice site are transferred to another spin state. For the $-\nicefrac{7}{2}$,$-\nicefrac{9}{2}$ mixture, the atoms in the $-\nicefrac{7}{2}$ are transferred to the $-\nicefrac{5}{2}$ state, while for the $-\nicefrac{5}{2}$,$-\nicefrac{9}{2}$ mixture, atoms in the $-\nicefrac{5}{2}$ are transferred to the $-\nicefrac{7}{2}$ state. These Zeeman-sublevels are then detected with the previously described Stern-Gerlach method. \\ 

\textit{Spin-correlations detection.} This measurement scheme is described in detail in Ref.~\cite{greif_short-range_2013}. In this paper, we only use the $-\nicefrac{7}{2}$,$-\nicefrac{9}{2}$ mixture in the spin-correlations measurement. As for the double occupancy detection, we start by ramping the lattice to a cubic 'freeze' lattice. We then eliminate the double occupancies: through consecutive radio-frequency Landau-Zener sweeps we transfer the atoms in the $-\nicefrac{7}{2}$ to the $-\nicefrac{3}{2}$ spin state. In doubly-occupied sites, the $-\nicefrac{3}{2}$, $-\nicefrac{9}{2}$ mixture is very short-lived, and therefore lost from the trap. The remaining single atoms in the $-\nicefrac{3}{2}$ state are transferred back to the $-\nicefrac{7}{2}$ state with two consecutive radio-frequency Landau-Zener sweeps.
The nearest-neighbour spin correlator along $t_x$ can be written as
\begin{eqnarray}
  \text{spin correlator} &=& -\frac{1}{N}\sum_{\vec{i}}\left[\langle\hat{S}_{\vec{i}}^x\hat{S}^x_{\vec{i}+\vec{e}_x}\rangle + \langle\hat{S}_{\vec{i}}^y\hat{S}^y_{\vec{i}+\vec{e}_x}\rangle\right]\quad\\
  &=& \frac{\text{singlet fraction} - \text{triplet fraction}}{2}\quad\text{,}\nonumber
\end{eqnarray}
where $\hat{S}_i$ is the standard vector spin-operator on lattice site $\vec{i}$, the sum runs over all two-site unit cells, $\vec{e}_x$ is the unit vector in $x$ direction, and $N$ is the total number of atoms in the ground band.
This means full antiferromagnetic correlations along $t_x$ would result in a spin correlator of one.
In order to measure the singlet and triplet fraction, we apply a magnetic field gradient which leads to an oscillation between the two populations. We measure these populations at the two extrema of the oscillations ($\SI{4.2}{ms}$ and $\SI{7.8}{ms}$) and infer the spin correlations from the difference. To do so, we ramp the lattice to a chequerboard configuration $V_{\overline{X},X,\widetilde{Y},Z}=\left[0.0,29.6(5),39.9(2),29.6(8)\right]$ thereby merging adjacent sites with singlets and triplets. 
Due to the Pauli exclusion principle, the triplets are then converted to one atom in the lowest and another in the first excited band. The singlets on the other hand form double occupancies in the lowest band. These single or double occupancies in the lowest band can be detected with the same method as described previously for double occupancies. To normalize these fractions we determine the total number of atoms $N$ including double occupancies in a separate measurement.

\subsection{Lifetime and band population fits}
To extract a characteristic timescale for the band population transfer and lifetimes for atom number, double occupancy fraction, and spin correlations we use an exponential decay function:
\begin{equation}
P\left(\tau\right)=P_0\exp\left(-\tau/\tau_P\right)\text{,}
\label{expDecay}
\end{equation}
where $P_0$ corresponds to the initial value of the observable, $\tau$ is the modulation time, and $\tau_P$ the fitted lifetime of the observable. 
The data points and corresponding error bars as a function of modulation duration correspond to the mean value and standard error from three (Fig.~3a-b, Fig.~4a) to ten (Fig.~4c) independent measurements. 
The curves in these Figures correspond to the result of fitting Eq.$~$\ref{expDecay} to all measured values.\\
To obtain an estimate on the uncertainty of the lifetime, we use two different sampling methods.
For the double occupancy fraction, atom number and ground band population (Fig.~3 and Fig.~4a-b) we use a method called bootstrapping: we randomly resample values from the measured points and fit the exponential decay function to this data set.
For the spin correlations, we assume a normal distribution around the mean measured value for each modulation time.
We then sample random values from these distributions and apply an exponential fit to the resulting data set.
We repeat both methods 500 times while varying the initialization parameters by plus minus 10$\%$: $P_0$ and $\tau_P$ for the atom number and band population and only $\tau_P$ for the double occupancy fraction and spin correlations ($P_0$ is then fixed to the initial value from the fit on all measured data points). 
We plot the mean standard deviation of the distribution of fitted parameters as shaded area in Fig.$~$3a-b and Fig.$~$4a,c and as asymmetric error bars in Fig.$~$3c-d and Fig.$~$4b.

\subsection{Band transfer calculations}
We calculate the band structure of our lattice and use the results to determine the expected population transfer between bands induced by the drive. We extend the methods described in Ref.~\cite{strater_interband_2016} to 2D lattices. In brief, we start with the time-dependent Hamiltonian where we include the modulation as a periodic relative modulation of the lattice depth:
\begin{equation}
\mathcal{H}=\frac{\mathbf{p}^2}{2m}+V\left(x,y,z,\tau\right)\text{,}
\end{equation}
where the lattice potential is similar to Eq.$~$\ref{Lattice}. The modulation is implemented by multiplying the non-interfering lattice depth in $x$-direction  with the time-dependent waveform of the modulation:
\begin{equation}
V_{\overline{X}}\left(\tau\right)=V_{\overline{X}}\times \left( 1+A_{\omega}\sin(\omega \tau)+A_{2\omega}\sin(2\omega \tau +\varphi)\right)
\text{.}
\end{equation}
The modulation in $x$-direction predominantly affects the dynamics in the $xz$-plane and we therefore neglect the $y$-direction in all subsequent calculations. To simplify further, we apply a coordinate transformation by rotating our frame by $\SI{45}{\degree}$:

\begin{align} 
x'=\frac{x+z}{\sqrt{2}} \qquad z'= \frac{z-x}{\sqrt{2}} 
\end{align}

The Bloch waves of the static Hamiltonian can be decomposed into their Fourier components:
\begin{equation}
\Psi_{q_{x'},q_{z'}}^n(x',z')=\sum\limits_{l,m=-\infty}^{+\infty}c_{l,m}^n(q_{x'},q_{z'})e^{i\left(\left(2lk_{L}+q_{x'}\right)x'+\left(2mk_{L}+q_{z'}\right)z'\right)}\text{,}
\end{equation}
Expressing the Hamiltonian in the basis of these coefficients $c_{l,m}^n(q_{x'},q_{z'})$ we can rewrite it to:
\begin{widetext}
\begin{equation}
\mathcal{H}_{l,l',m,m'}=\left\{
\begin{array}{ll}

\vspace{1mm}		
\frac{\hbar^2}{2m}\left(2\left(\frac{q_{x'}}{2}+ l k_{L}\right)^2+2\left(\frac{q_{z'}}{2}+ m k_{L}\right)^2\right) & \text{for } l=l' \text{ and } m=m'\\
\vspace{1mm}		

-\frac{\alpha}{2}\sqrt{V_{X}V_{Z}}\cos(\phi)  & \text{for } \left(|l-l'|=1 \text{ and }m=m'\right) \text{ or } \left(l=l' \text{ and }|m-m'|=1\right)\\

\vspace{1mm}		
-\frac{V_Z}{4} &\text{for } l=l'\pm 1\text{ and } m=m'\pm 1\\

\vspace{1mm}		
-\frac{V_{\overline{X}}(\tau)}{4}-\frac{V_{X}}{4}e^{\pm i\theta}
& \text{for }l=l'\pm 1 \text{ and } m=m'\mp 1\\

0 & \text{otherwise.}

\end{array} 
\right.
\end{equation}
\end{widetext}
We sample the positive quadrant of $(q_{x},q_{z})$ in steps of $0.25 \times k_L$. For each $q$-vector we integrate the time-dependent Schroedinger equation for a finite number of evenly spaced steps in time up to a time $\Delta \tau$ which corresponds to the modulation time.
We assume for the atoms to be initially in the lowest band and evolve that state in time. For each time step, we calculate the overlap of the time-evolved state with the static eigenstate of the ground band. 
The curves in Fig.$~$2a and Fig.$~$3 are obtained by taking the minimum overlap for each quasi-momentum and averaging over the entire Brillouin zone.\\

The Hubbard parameters $t$ and $U$ are numerically calculated from the Wannier functions of the lattice potential, which we obtain from band-projected position operators~\cite{uehlinger_artificial_2013}.
\clearpage
\end{document}